\documentclass[manuscript,screen]{acmart}
\definecolor{teal}{RGB}{0,183,150}

\usepackage{bm}

\setcopyright{acmcopyright}
\copyrightyear{2022}
\acmYear{2022}
\acmDOI{XXX}

\acmConference[CHI'22]{CHI TRAIT '22: Workshop on Trust and Reliance in AI-Human Teams}{2022}{New Orleans, LA}
\acmBooktitle{CHI TRAIT '22: Workshop on Trust and Reliance in AI-Human Teams,
  2022, New Orleans, LA}
\acmPrice{XXX}
\acmISBN{XXX}

\begin{document}

\title{The Value of Measuring Trust in AI -- A Socio-Technical System Perspective}


\author{Michaela Benk}
\affiliation{%
 \institution{ETH Zurich}
 \country{Switzerland}}

\author{Suzanne Tolmeijer}
\affiliation{%
 \institution{University of Zurich}
 \country{Switzerland}}
 
\author{Florian von Wangenheim}
\affiliation{%
 \institution{ETH Zurich}
 \country{Switzerland}}

\author{Andrea Ferrario}
\affiliation{%
 \institution{ETH Zurich}
 \country{Switzerland}}

\renewcommand{\shortauthors}{Benk, et al.}

\begin{abstract}
Building trust in AI-based systems is deemed critical for their adoption and appropriate use. Recent research has thus attempted to evaluate how various attributes of these systems affect user trust. However, limitations regarding the definition and measurement of trust in AI have hampered progress in the field, leading to results that are inconsistent or difficult to compare. In this work, we provide an overview of the main limitations in defining and measuring trust in AI. We focus on the attempt of giving trust in AI a numerical value and its utility in informing the design of real-world human-AI interactions. Taking a socio-technical system perspective on AI, we explore two distinct approaches to tackle these challenges. We provide actionable recommendations on how these approaches can be implemented in practice and inform the design of human-AI interactions.
We thereby aim to provide a starting point for researchers and designers to re-evaluate the current focus on trust in AI, improving the alignment between what empirical research paradigms may offer and the expectations of real-world human-AI interactions.
\end{abstract}




\maketitle

\section{Introduction}



Artificial Intelligence (AI)-based systems\footnote{In this work, we focus our attention on those AIs using machine learning models only. We therefore consider ``AI'' or ``AI system'' as a synonym for ``machine learning-based system''.} are supporting humans in an increasing number of decisions and in a multitude of society-impacting applications \cite{venkatasubramanian2020philosophical, Chiou2021TrustingAD}. As building trust in such systems is deemed critical for their adoption and appropriate use \cite{Knowles2021TheSO,hurlburt2017much}, much attention in recent research in computer science, human-computer interactions (HCI), and explainable AI (XAI) has been given to the development of trustworthy AI guidelines \cite{toreini_relationship_2019, Jobin-et-al_2019}, and methods to foster and evaluate trust in human-AI interactions \cite{Ribeiro-et-al_2016,Vereschak2021HowTE, Jacovi2020FormalizingTI}.

To this end, various studies investigate the relation between understanding the ``black box model'' and the ability to calibrate trust in it (e.g., \cite{Langer2021WhatDW,Lakkaraju2017InterpretableE,wanner2020white}). The working hypothesis suggests that by providing adequate methods to increase understanding of machine learning (ML) models, i.e., via explanations, trust in the models, their appropriate use \cite{Arrieta2019ExplainableAI, gilpin-et-al_2018} and adoption \cite{Knowles2021TheSO} would follow. In contrast, an excess (or lack) of trust may result in the misuse, disuse, or abuse \cite{Glikson2020HumanTI, Hoff-Bashir_2015, Jacovi2020FormalizingTI,lee_trust_2004} of the models.

Results of studies aiming to test the hypothesis have shown to be inconsistent: users at times over-trust models or model explanations \cite{kaur-et-al_2020, Bansal2021DoesTW, PoursabziSangdeh2021ManipulatingAM, Lakkaraju2020HowDI}, while at other times they inappropriately distrust the AI \cite{Yang2020HowDV, stumpf-et-al_2016}. Additionally, some studies report mixed results, where explanations might increase understanding, but not trust \cite{Chiou2021TrustingAD, Kastner2021OnTR, Ehsan2021ExpandingET}. These inconsistencies may hamper progress in the field and hinder the successful deployment and use of AI-based systems in practice. For instance, at the time of writing, XAI methods fall short of assisting heterogeneous stakeholders beyond data scientists or engineers and beyond the purpose of debugging or validating ML models \cite{Bhatt2020ExplainableML}.

Recent works have related these inconsistencies to current empirical practices aiming at defining and evaluating trust in AI \cite{Jacovi2020FormalizingTI, toreini_relationship_2019, Chita-Tegmark-et-al_2021, Buccinca2020ProxyTA, Kastner2021OnTR, Chiou2021TrustingAD}. First, numerous terms have been introduced to indicate trust and trust-related concepts in empirical studies, including ``calibrated trust'' \cite{Langer2021WhatDW}, ``appropriate trust'' \cite{Yang2020HowDV}, ``reliance'' \cite{hoffman2018explaining}, and ``actual trustworthiness'' \cite{Schlicker2021TowardsWT}. This has resulted in extensive theoretical discussion on what is meant by the construct ``trust in AI'' or what it should mean \cite{Lewis2022WhatII, Deley2020AssessingTV,hatherley2020limits,ferrario2021trust,Luhmann2000FamiliarityCT, Siegrist_risk-confidence2005,Jacovi2020FormalizingTI,loi2020much}. A unified language is necessary to compare results and work toward a common research stream.  

Second, the empirical setting in which trust in AI is measured is often limited --- it includes simplified laboratory studies with proxy tasks \cite{Buccinca2020ProxyTA}, does not incorporate the risk factor of trust \cite{Jacovi2020FormalizingTI}, focuses on humans in supervisory control roles over other settings \cite{Chiou2021TrustingAD}, and promotes measurement with self-reported questionnaires that do not fit the context \cite{Chita-Tegmark-et-al_2021}, or behavioral measures that do not match self-reported results \cite{Buccinca2020ProxyTA, Zhang-et-al_2020, Papenmeier2019HowMA}. Crucially, at the basis of these experiments lies the attempt to encode the construct ``trust in AI'' in a single numerical value that, arguably, should appropriately inform the design of AI systems for real-world applications.

In this work, we contribute to the discussions on the measurement of trust in AI as follows. First, we briefly outline the main limitations in defining and measuring trust in AI by discussing examples of empirical studies in XAI. In particular, we highlight the challenge of attributing a meaning to the numerical value of trust in AI and using it to inform the design of real-world applications. Second, we propose to view AIs through the lens of socio-technical systems \cite{van2020embedding} as an attempt to better align the discourse of trust in AI to the expectations and specificities of human-AI interactions in real-world settings. In doing so, we discuss two distinct approaches to measuring trust in AI. The first--constructivist--approach aims to measure trust in AI as the aggregation of trust levels in all interactions of the socio-technical systems. The second--paradigm shifting--approach suggests to refocus research efforts away from \textit{a priori} definition and measurement of trust in AI. Instead, we propose to consider trust-enabling constructs, such as explanation's utility, critiquing, and team cognition, that may allow designers to evaluate and improve the quality of the interactions of a socio-technical system. 


\section{Measuring Trust in AI and Assessing its Value}
 \label{valuevalue} 
 

\subsection{Empirical measurement of trust -- current approaches}\label{section:trust_mesure}
Despite the large interdisciplinary interest in trust in AI, there is still no consensus on a clear definition of the construct, how it relates to trust in human interactions or other fundamental constructs, such as user understanding or mental models. Moreover, the measurement of trust in AI and the relationship between different types of measures of it remain open research issues. Arguably, this is not surprising, as the same limitation holds for trust in interpersonal relations (see., e.g., \cite{McLeod2006Feb,McEvily_tortoriello_2011}).
Nevertheless, different accounts of trust in AI are currently under investigation in the XAI and HCI domains, fueling the design of many empirical studies (e.g., \cite{Jacovi2020FormalizingTI, toreini_relationship_2019, Glikson2020HumanTI, Hoffman2018MetricsFE, hatherley2020limits}). On the one hand, researchers' use of terminology and trust definitions is diverse: commonly used terms include ``calibrated trust''\cite{Tenhundfeld2019CalibratingTI}, ``contractual'' and ``warranted trust'' \cite{Jacovi2020FormalizingTI}, ``appropriate trust'' \cite{hoffman2018explaining}, ``reliance''\cite{nourani2021anchoring}, and ``perceived trustworthiness''\cite{Weitz2020LetME}, to name a few. The lack of a uniform language makes it difficult to establish common ground and to ensure that the field progresses in a unified research stream.

On the other hand, empirical studies that aim at measuring trust in AI often share common design patterns and limitations, which become particularly evident in the case of studies on trust in the context of XAI \cite{Buccinca2020ProxyTA, Jacovi2020FormalizingTI}. For instance, while experimental studies should follow the steps of 1) statement of the problem, including hypotheses to test, 2) selection of the response variable(s), and 3) choice of factors, levels, and ranges \cite{montgomery2017design,booth2003craft}, these studies often fall short in fully explicating their hypotheses. One example is the ``running hypothesis'' on trust in AI \cite{miller_explanation_2018}, i.e., the idea that increasing understanding in a ML model would foster trust in it, which is implemented in different ways in different studies (e.g., as separate ``trust'' and ``understanding'' outcome variables \cite{kaur-et-al_2020} or only via self-reported and behavioral trust measures \cite{Yin2019UnderstandingTE}). Furthermore, studies measuring trust in XAI often implement simplified tasks, such as validating a sequence of pre-computed model outcomes, to be performed in laboratory settings (i.e., simulated in an artificial setting) by human users with fixed or limited types of roles, such as that of a  human in the supervisory role and AI as subordinate \cite{Chiou2021TrustingAD}. Often these roles are filled by MTurkers \cite{PoursabziSangdeh2021ManipulatingAM, Buccinca2020ProxyTA, Bansal2021DoesTW, Yin2019UnderstandingTE} or ML experts, such as engineers or data scientists \cite{kaur-et-al_2020, Zhang-et-al_2020}. 
Moreover, experimental protocols do not explicitly encode key components of trusting relationships, such as risk \cite{Jacovi2020FormalizingTI}, and do not capture the dynamics of trust, i.e., its changing over time with user's increasing experience and skills, and different roles \cite{Chiou2021TrustingAD}. Finally, in these studies, the measurement of trust is performed with a myriad of different instruments. Examples are self-assessment questionnaires \cite{Cheng-et-al_2019,Lucic2020WhyDM,Shin2021TheEO,Ribeiro-et-al_2016,Yin2019UnderstandingTE}, 
behavioural measures (e.g., counts of agreements with the ML model in a sequence of decisions) \cite{kim2021should,scharowski2021initial,Nourani2020TheRO}, or qualitative measures, based, for instance, on interview transcripts \cite{Buccinca2020ProxyTA,brennen2020people,danry2020wearable,ehsan2019automated}.

In summary, lack of consensus on the terminology used around trust in AI, limitations in experimental protocols of studies aiming to evaluate trust--including those pertaining the use of instruments to measure the construct--significantly affect current research on human-AI interactions. They impede the comparability and generalizability of empirical results. 
One could argue that, despite the above limitations affecting the definition and measurement of trust in AI, we could continue measuring trust because of its presumed utility in supporting the design of appropriate human-AI interactions. We discuss this point in the forthcoming section, by analyzing the intrinsic value of its measurement and its use by designers of human-AI interactions.

\subsection{The value of giving trust in AI a value}\label{section:trust_value}
Underlying experiments measuring trust in AI is the idea that trust can be encoded in a numerical value\footnote{In this section--including its title--, we will use the term ``value'' as ``relative worth, utility, or importance'' \cite{merriam_webster} of something, or the ``numerical quantity that is assigned or is determined by [...] measurement'' \cite{merriam_webster} of a quantifiable variable. In the latter sense, we will sometimes use the expression ``numerical value'' for the sake of clarity.} that is computed using different instruments and methods, as discussed in Section \ref{section:trust_mesure}.


Currently, empirical studies to measure trust in AI collect only a few measurements of the construct for each participant in single-session experiments. Typically, each participant answers a questionnaire once per session, or a behavioural measure of trust in AI is collected at each step of a given sequence of cases (e.g., machine learning models outcomes to be validated by each participant). Then, the measurements for each participant are aggregated at the level of each experimental condition by computing, for example, the mean of their distribution. 
There are two open challenges in this approach. First, we are left with the attempt to understand, for example, what a measurement of 5 on a 7-point Likert-scale may mean and how it differs from a measurement, let us say, of 6 on the same scale. Second, we would need to understand how the difference between 6 and 5 may relate to the one between 5 and 4 using the same (or a different) scale, in the same (or in a different) context.
In fact, without a quantitative model of trust in AI, i.e., a function relating levels of trust to context-dependent variables that are deemed relevant in human-AI interactions (e.g., risk, objective trustworthiness of the AI), it is not possible to provide an interpretation of these values and their differences beyond the comparison of their absolute values. Finally, we would need to collect measurements across multiple sessions, both at the participant and experimental condition levels.

On the one hand, the literature on computational trust and logic \cite{sabater2005review,cho2015survey,marsh2005trust,primiero2012modal}, and trust in human-AI interactions \cite{taddeo2010modelling,loi2020much,ferrario_ai_2019} offers a few examples of quantitative models of trust. In particular, these latter examples propose to measure trust as a function of measurable quantities in the interaction, such as the amount of planned resources (or monitoring) to be invested in it and a quantification of the trustor's stakes. However, to the best of our knowledge, none of the models have been yet applied in HCI studies to measure trust in AI. Therefore, their applicability is yet to be tested. On the other hand, calibrating trust with trustworthiness \cite{lee_trust_2004,Jacovi2020FormalizingTI} may provide a support at interpreting values of trust in AI. Jacovi et al.'s recent account of contractual trust \cite{Jacovi2020FormalizingTI}, i.e., the belief that and AI will maintain a functionality--called ``contract''--deemed relevant in a given context, has further fueled the interest in trust calibration. However, the applicability of calibration in empirical studies is conditional on the 1) selection of a set of contracts in a given context, 2) the appropriate quantification of (continuous) levels of objective trustworthiness of the AI in the selected contracts, and 3) the measurement of trust levels \cite{lee_trust_2004}. Only then, the comparison of trust levels (e.g., the mean level of each experimental condition) may provide support to the analysis of trust in the context of calibration. However, the literature on human-AI interactions has not yet arrived at clear recommendations and a structured approach to trust calibration in empirical studies, beyond the case of straight-forward AI contracts, such as stated model accuracy or prediction confidence scores \cite{Zhang-et-al_2020, Yin2019UnderstandingTE}. 

Finally, some might argue that trust experiments in AI are comparative, i.e., they aim to compare treatments \cite{cox1958planning}, such as different explainability methods or levels of trustworthiness in maintaining a contract, rather than to compare values of trust in AI \textit{per se}. However, we argue, without an understanding of the values of the construct at hand and a consistent measurement procedure, not even the understanding of uncontrolled variations in each experimental condition \cite{cox1958planning}, let alone the comparison of different treatments or even experiments, can be possible. To adapt the example by Chiou and Lee, how can we compare the same value of trust in an autonomous car for two scenarios with significantly different outcomes (e.g., one resulting in a delay, versus one resulting in the death of a passenger) \cite{Chiou2021TrustingAD}? \\

\textit{Measuring trust in AI to inform the design of AI systems.} Finally, it is not clear how a numerical trust value may inform the development of real-world applications where humans and AIs interact. For example, the comparison of mean values of trust across empirical conditions may indicate that an explainability method is perceived as more trustworthy than another. However, what actionable recommendations for designers could possibly follow from this fact? On the one hand, they could be interested only in the comparison between experimental conditions \cite{cox1958planning} and promote the implementation of the explainability method with highest mean trust value. On the other hand, they could consider also constraints, such as a minimum (average) value for trust, to support the subsequent design of the system. In this case, the constraints could be the result of considerations that may follow from the analysis of comparable HCI studies, the measurement of additional outcome variables (such as user experience, usability, or participant's performance), 
and pragmatic reasons, such as management heuristics or corporate requirements for the deployment of trustworthy AIs. As a result, it is not clear which approach, or set of guidelines, could assist the transfer of the results of the measurement of trust in AI into the design, development, and deployment of AIs for real-world applications. At best, the measurement of trust in AI (however it is operationalized)  provides only one of the many inputs that are necessary to plan the design of AI systems and their real-world applications. This inevitably limits the impact of the studies on trust in AI that focus exclusively on lab experiments and do not tackle on the challenge of a trust-informed design and testing of AIs. This limitation is particularly relevant in healthcare, as some authors have highlighted the necessity of defining a ``design science for AI'' to assist the design and development of trustworthy medical AI systems \cite{yang2019unremarkable}.

\section{The different types of trust in human-AI interactions: a socio-technical system perspective}\label{section:socio}
As mentioned in Section \ref{section:trust_mesure}, empirical studies aiming to measure trust in human-AI interactions often consider simplified laboratory settings. In fact, in these interactions, the human agent often acts as a supervisor of the AI, which, in turn, provides a sequence of tasks for the human to undertake; above all, the analysis of ML model outcomes. Neither the human agent, nor the AI interact with other stakeholders or AIs, and they keep their assigned roles throughout their relation. Arguably, this design may have been influenced by the literature on human-machine interactions in the domain of automation,\footnote{For example, an often used trust questionnaire by \citet{Jian2000FoundationsFA} is, in fact, developed to measure trust in automation rather than AI.}
where mostly repetitive tasks are performed in a traditional supervisory control setting \cite{Sheridan1978HumanAC, Miller2007DesigningFF}. 
The key observation is that the supervisory control setting fails short of appropriately representing the heterogeneous human-AI interactions that take place in real-world applications. Therefore, although it may still be adequate to represent specific interactions (for example those where data scientists performing debugging and validation of an AI), we propose a change of perspective in how trust in AI should be related to real-world situations where human agents and AIs interact. To do so, we propose to consider AIs as socio-technical systems, i.e., ``systems that depend on not only technical hardware but also human behavior and social institutions 
for their proper functioning'' \cite{van2020embedding}, where social institutions are the rules of the interactions between humans and systems. We discuss this proposition in the following paragraphs. 

\subsection{AIs as socio-technical systems}
By definition, in a socio-technical system, multiple stakeholders, e.g., engineers, business owners, and customers interact with the AI to achieve their goals according to their roles, through the use of system interfaces, over time, and following the rules of the system social institution \cite{van2020embedding,kroes2006treating}. Therefore, the definition of an AI as a socio-technical system requires the specification of the stakeholders, their interactions with the AI, and the set of rules in which these interactions take place. 
Originally, the idea of considering AIs as socio-technical systems stems from recent literature on the ethics of AI \cite{van2020embedding,theodorou2020towards,loi2021transparency}.
However, it has not yet been specifically adopted for the measurement of trust in AI and to try to overcome the limitations of the supervisory control setting and its empirical paradigms. 
A socio-technical system perspective of human-AI interactions can enrich the current approach to trust in AI and its measurement by providing a basis for a discourse on the definition, implementation, and control of values in trustworthy AI systems \cite{van2020embedding}, such as the respect for autonomy of its users, fairness, transparency, and explainability \cite{hleg2019ethics}. Therefore, this perspective becomes especially relevant, we argue, to try to align the research methods in human-AI interactions with the expectations of real-world applications. 

Let us clarify our proposal by introducing an example of an AI as a socio-technical system from healthcare. Consider a medical AI system deployed in an orthopaedic clinic to predict whether radiographs may display small fractures in human bones. Examples of these fractures are scaphoid breaks that are notably hard to detect on medical images \cite{yoon2021development}. Such an AI can improve clinicians' work by supporting the timely identification of fractures and administration of appropriate treatments. The socio-technical system at hand comprises human stakeholders, such as patients, orthopaedicians (medical students, residents, and orthopaedic surgeons), data engineers, data scientists, and forensic pathologists interacting with the AI in the orthopaedic clinic. Specifically, the latter defines the social institutions of the system. Human-AI interactions in the system may comprise the periodic validation of batches of images, AI outcomes and the performance of the AI (data engineer, data scientist, clinicians, and the AI), the discussion of a given image and its AI outcome to determine diagnosis (patient, clinician, and the AI), or the treatment for the fracture (clinician,  surgeon, and the AI). 
Moreover, the AI may change role within the socio-technical system, as the result, for example, of a change in its integration in the decision-making processes in the orthopaedic clinic \cite{yu2018artificial}. This would result in the change (including the cessation) of some interactions of the socio-technical system, due to, for example, an update of the roles of the stakeholders interacting with the newly-integrated AI. Motivated by these discussions, we arrive at the key questions: 
\begin{quote}
\textbf{Given the proposal of considering a socio-technical system perspective in human-AI interactions, what is the ``trust in AI'' to be considered? Should we consider measuring it? If yes, how?}    
\end{quote}

We answer them by introducing two approaches that, we hope, can inform future research on trust in AI.

\subsection{A constructivist approach to trust in AI and its measurement in socio-technical systems}\label{section:constructive}
\begin{figure}[ht]
 \centering
  \includegraphics[width=0.6\linewidth]{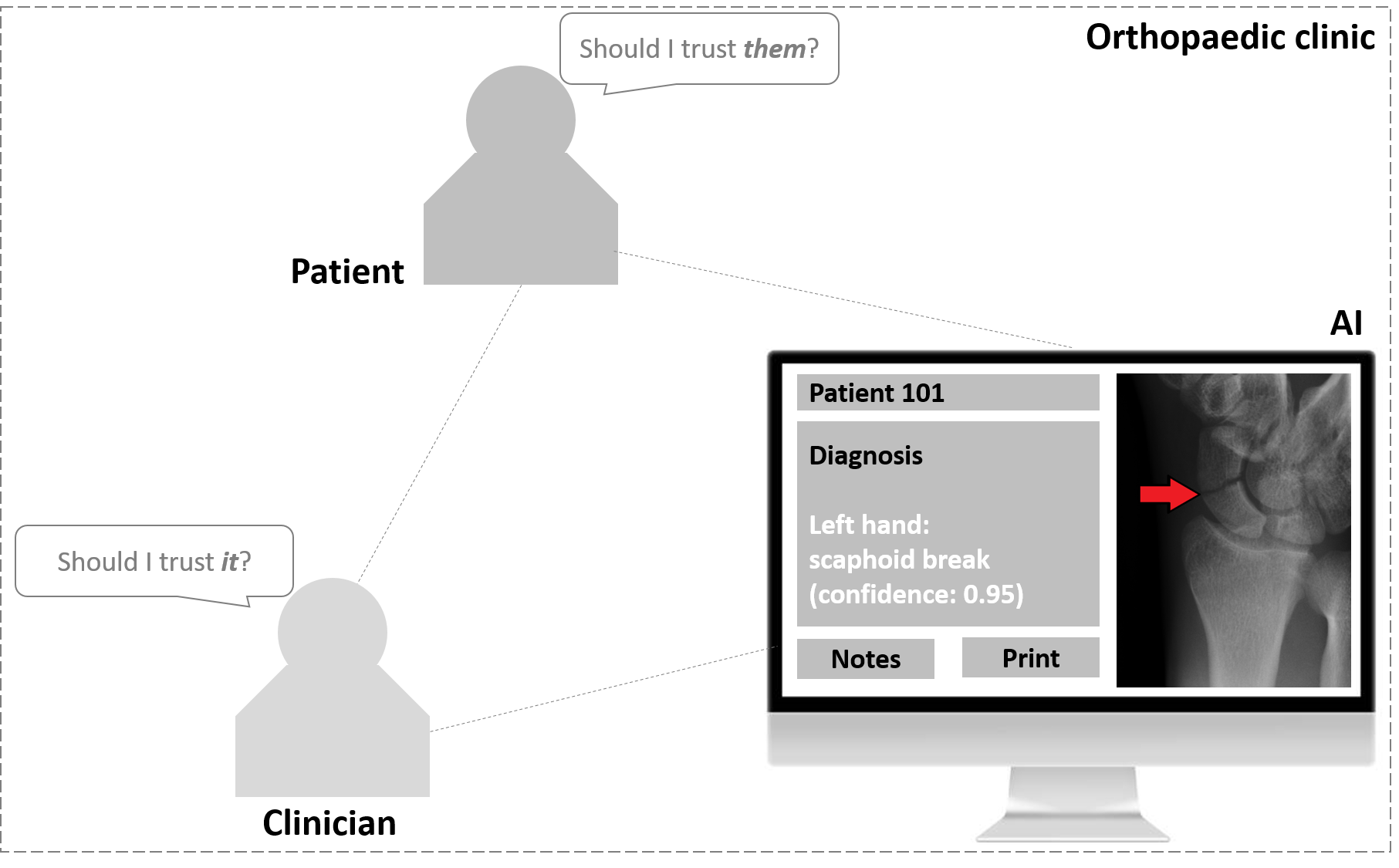}
  \caption{An interaction in the socio-technical system with two human stakeholders and the AI. Given the interaction, we can measure trust in AI by considering the clinician and the AI, and the trust of the patient in the dyad clinician-AI. The different stakeholder perspectives and goals go beyond the possibilities offered by supervisor control settings and their measurements of trust in AI}
  \label{fig:three_relation}
\end{figure}

As a first possibility, we could embrace the perspective of an AI as a socio-technical system by modeling trust in it at the \textit{system} level. In doing so, we do introduce the ``trust in AI'' to be considered by aggregating all AI-relevant interactions in the system and specifying a constructivist procedure to measure it. To follow this approach, designers need to:
\begin{enumerate}
    \item specify the elements of the socio-technical system, i.e., the AI as technical artifact and its requirements, the stakeholders, their roles, and activities, and the applicable social institutions  \cite{paja2013trust},
    \item formally define the interactions involving stakeholders and the AI that are deemed relevant for analyzing and measuring trust in AI, such as by creating a rooted graph of all possible stakeholder interactions.
    \item take care of the dynamics of the system, by 
    identifying a finite sequence of snapshots of the system at different times (as the set of interactions in the system may evolve over time),
    \item measure trust for each interaction in the system, at each time $t$, 
    \item compute the value of trust in the system at each time $t$ by aggregating all the contributions from (4), e.g. by using a weighted mean \cite{grabisch2009aggregation},  
    and 
    \item compute the value of trust in the socio-technical system by aggregating all snapshot contributions, as in (5).
\end{enumerate}
At the basis of the constructivist approach lies the idea that trust in AI---when an AI is seen through the socio-technical system lens---can be computed via a function that aggregates all the trust measurements from the different interactions of the system over time. The value of considering an AI as a socio-technical system level is to provide a context-aware perspective to trust in AI that considers \textit{all} the contributions of the stakeholders interacting with the AI, including their dynamics. 
As a result, this approach may better encode the real-world specificities of the use of an AI than existing empirical paradigms, such as those promoting the supervisory control setting and in simplified laboratory 
human-AI interactions. It may also provide actionable support to designers aiming to deploy AI systems in real-world applications by  identifying and intervening on those interactions that results in low levels of trust as compared to the others in the system, over a certain period of time. Clearly, to implement the approach designers would need to
\begin{enumerate}
\item measure trust in the interactions of the system, and 
\item introduce mathematical rules (and their justification) to aggregate the different measurements of trust.
\end{enumerate}
To do so, we suggest the use of the same method for all interactions in the system, and relying on quantitative models of trust \cite{cho2015survey,marsh1994formalising,taddeo2010modelling,loi2020much}. However, we note that existing models of trust  \cite{taddeo2010modelling,loi2020much} do not consider the case of relations where two or more stakeholders interact with an AI.
In fact, trust is traditionally modeled as a three-placed relation between a trustor $X$, a trustee $Y$ and a goal $g$. Therefore, the measurement of trust in the case of human-AI interactions involving multiple stakeholders is an open research problem asking for appropriate formalization. For example, how to compute the construct ``trust in AI'' in a relation during which a patient has a consultation with an orthopedist and together they discuss the outcome of an AI that has identified a small fracture in the patient's radiography? In this case, \textit{who trusts whom (or what)?} We show this example in Figure \ref{fig:three_relation}. To address this multi-stakeholder perspective in human-AI interactions, designers may consider the aggregation of the measurements of trust over the relevant dyads in the interaction \cite{ferrario2022explainability}, i.e., clinician-AI, patient-AI, (clinician-AI)-patient etc. However, this and other proposals, such as those focusing on team cognition and trust from the Computer-Supported Cooperative Work domain \cite{Schelble2022LetsTT} still need to be thoroughly tested in empirical studies. 


\subsection{Toward trust enablement}

An alternative and, perhaps, more provocative approach to the definition and measurement of trust in AI may simply suggest to stop trying to define and measure it. Followers of this proposals may argue that the severity of the limitations highlighted in Section 2.1 and the complexity of an AI as a socio-technical system justify the refocus of research efforts on constructs that 1) may be more explicit in their definitions, 2) can be consistently measured, and 3) can appropriately inform AI system design for real-world applications.

Endorsing this \textit{paradigm shifting} approach, we aim to empower users by creating the conditions for letting them interact with AI in real-world applications at their best \cite{Dwyer2011TracesOD, Marsh2020ThinkingAT}. Pragmatically, this means to minimize the risks stemming from the use of the AI and the mistakes resulting from relying on it, providing guidance in case of conflicts with the system, fostering learning over time from its repeated use (possibly in a team setting), using AI inputs for decision-making in existing workflows and in a timely fashion, and fostering actionable recourse against unfavourable AI outcomes. We argue that striving for these pragmatic goals may be related to what is commonly described as ``trust in AI'', or what is reasonable to expect in an interaction with a trustworthy AI, whatever its precise definition and multi-dimensional, context-specific measurement may possibly be. However, following this approach, none of the two need to be explicitly introduced: only the (possibly) ``trust-enabling'' constructs to achieve the pragmatic goals do. As a result, this trust enablement paradigm, ``allow[s] users to define what trust means to them in a certain situation, and to allow a user to conduct a trust relationship that is in their best interests and on their own terms'' \cite{Dwyer2011TracesOD}. In summary, it is not a numerical value of trust in AI that defines the quality of an interaction within the socio-technical system, but, rather, the result of evaluating a set of actionable constructs that encode different characteristics and—reasonable, from a pragmatic perspective--expectations of the interaction itself.

The operationalization of the proposed trust enablement paradigm for AI systems remains, to the best of our knowledge, an open research challenge. One way to introduce constructs for trust enablement is to focus on the elements in each interaction within the socio-technical system that allow for improved understanding, learning, or critical assessment of AIs and their outcomes, in single or multi-user contexts. For example, the measurement of the change in the utility of an explanation of an AI, or the explanation of an AI outcome provided by another stakeholder (as in the clinician-patient-AI interaction, see Figure \ref{fig:three_relation}) \cite{Davis2020MeasureUG, Hsiao2021RoadmapOD}, may provide a realistic evaluation of users' attitudes toward the AI and insights for individual requirements over time, across different roles, interactions, and tasks in the system.
To this end, as real-world interactions with AIs become increasingly interactive, novel techniques, such as critiquing \cite{antognini2020interacting}, may help improving the interactions and represent a measurable construct for trust enablement in a socio-technical system. Critiquing may enable the interactive recourse against an AI outcome, such as the prediction of a small fracture, through the exchange of information with the system and the implementation of user feedback. For example, a patient may critique the dyad clinician-AI by challenging their prediction of a scaphoid break on the basis of the absence of pain and swelling in his hand (see \ref{fig:three_relation}). The dyad clinician-AI would then provide a justification for the prediction by arguing that the size of the fracture is compatible with the absence of tissue swelling (as shown in the radiography) and localized pain. Showing markers on the radiography and retrieving analogous images from the electronic health records may support the successful finalization of the critiquing process. Measuring critiquing may result in evaluating the total number of interactions in the process, the time spend to agree on the AI outcome, and the change in the patient's confidence in the dyad competence levels, at the beginning and the end of the process. Finally, as the clinician mediates between the AI and the patient, we argue that understanding and characterizing the contribution of the clinician's support in the process of critiquing that sees the patient interacting with the clinician-AI dyad becomes a research priority for designers. 
In addition to critiquing, other constructs for trust enablement include common ground, which refers to agents’ mutual knowledge and understanding of a joint activity \cite{Clark1996UsingL}, negotiation, which requires clarifications or conflict resolutions \cite{Beers2006CommonGC}, and, finally, team cognition \cite{Schelble2022LetsTT}. This latter is an emerging research agenda in the Computer-Supported Cooperative Work domain. It encompasses also the case of human-AI teams, which is applicable, for example, to the clinician-AI dyad example \cite{Liu2021UnderstandingTE, Bansal2019BeyondAT}. Nevertheless, such works are still in their infancy. The use and implementation of team cognition and interactivity in socio-technical systems may be key in improving the quality of the socio-technical system interactions as it may allow, in particular, ``advancing real-world design recommendations that promote human-centered teaming agents and better integrate the two'' \cite{Schelble2022LetsTT}. We plan to explore this avenue of research in future work.

\section{Conclusion}

We have shown that the existing approaches to the definition and measurement of trust in AI are affected by limitations that impact, in particular, its use to inform the design of real-world applications. As a first attempt to tackle this challenge, we have proposed to consider AIs through the lens of socio-technical systems and to embrace the complexity of all possible interactions that humans may have with the AI. 
To do so, we proposed two actionable approaches to the measurement of trust in AI. The first suggests that it is possible to measure trust in AI in the socio-technical system perspective, if designers agree on the methods to measure it for all interactions of the system and the logic of aggregation of these measurements. The second proposes to focus on more pragmatic and better defined constructs that would enable trust in AI, allowing designers to evaluate the quality of the interactions in the socio-technical system, instead.  With these proposals, we hope to support empirical researchers in re-evaluating the current focus on trust in AI, inspiring a discussion on new approaches to its definition and measuring, and reducing the distance between what empirical research paradigms may offer and the expectations of real-world human-AI interactions.

\section{Authors Contributions}
MB and AF drafted the manuscript, with helpful support of ST, and valuable feedback of FW on the final version.

\bibliographystyle{ACM-Reference-Format}

\end{document}